\documentclass[journal]{IEEEtran}
\usepackage{amsmath, amssymb, bm, cite, epsfig, psfrag, mathtools, gensymb}
\usepackage{epstopdf}
\usepackage{rotating}
\usepackage{dblfloatfix}
\usepackage[margin=0.61in,top=0.881in]{geometry}
\usepackage{algorithm,algpseudocode}
\usepackage{array}
\usepackage{supertabular,booktabs}
\usepackage{ragged2e}
\newcolumntype{P}[1]{>{\centering\hspace{0pt}}p{#1}}
\newcolumntype{M}[1]{>{\centering\hspace{0pt}}m{#1}}
\newcolumntype{L}{>{\centering\arraybackslash}m{3cm}}
\usepackage{changepage}
\usepackage[font=small]{caption}
\usepackage{subcaption}
\usepackage{mwe}
\usepackage{color,soul}
\usepackage{multicol}
\makeatletter
\newcommand*{\balancecolsandclearpage}{%
  \close@column@grid
  \cleardoublepage
  \twocolumngrid
}

\usepackage{capt-of}
\usepackage{bbm}
\usepackage{multirow}
\usepackage[usenames,dvipsnames]{xcolor}
\renewcommand{\arraystretch}{1.5}

\usepackage{footnote}
\makesavenoteenv{tabular}
\makesavenoteenv{table}
\usepackage{etoolbox}
\usepackage{pbox}
\usepackage{cite}
\usepackage{longtable}
\usepackage{tabu}

\usepackage{pdflscape}
\usepackage{graphicx}
\usepackage{tabu}
\newcolumntype{?}{!{\vrule width 1.6pt}}
\usepackage{booktabs}
\usepackage{todo}
\usepackage{tikz}
\usetikzlibrary{calc}

\def\dB{\textrm{dB}}
\def\dBm{\textrm{dBm}}

\def\FS{\textrm{FS}}

\def\1m{\textrm{1 m}}

\begin{document}
\title{Indoor Office Wideband Penetration Loss Measurements at 73 GHz}

\author{
	\IEEEauthorblockN{Jacqueline Ryan,~\IEEEmembership{Student Member,~IEEE,} George R. MacCartney, Jr.,~\IEEEmembership{Student Member,~IEEE,}\\
		and Theodore S. Rappaport,~\IEEEmembership{Fellow,~IEEE}\\
		\IEEEauthorblockA{NYU WIRELESS and NYU Tandon School of Engineering, New York University, Brooklyn, NY 11201}\vspace{-0.7cm}
	}
	
	\thanks{This material is based upon work supported by the NYU WIRELESS Industrial Affiliates Program, three National Science Foundation (NSF) Research Grants: 1320472, 1302336, and 1555332, and the GAANN Fellowship Program. The authors thank Y. Xing, H. Yan, and S. Sun for their help in conducting the measurements. J. Ryan (email: jpr369@nyu.edu), G. R. MacCartney, Jr. (email: gmac@nyu.edu), and T. S. Rappaport (email: tsr@nyu.edu), are with the NYU WIRELESS Research Center, NYU Tandon School of Engineering, New York University, Brooklyn, NY 11201.}
}

\maketitle
\begin{tikzpicture}[remember picture, overlay]
\node at ($(current page.north) + (0,-0.4in)$) {J. Ryan, G. R. MacCartney, Jr., and T. S. Rappaport, ``Indoor Office Wideband Penetration Loss Measurements at 73 GHz,"};
\node at ($(current page.north) + (0,-0.55in)$) {in \textit{2017 IEEE International Conference on Communications Workshop (ICCW)}, Paris, France, May 2017, pp. 1-6.};
\end{tikzpicture}

\begin{abstract}
This paper presents millimeter wave (mmWave) penetration loss measurements and analysis at 73 GHz using a wideband sliding correlator channel sounder in an indoor office environment. Penetration loss was measured using a carefully controlled measurement setup for many common indoor building materials such as glass doors, glass windows, closet doors, steel doors, and whiteboard writing walls. Measurements were conducted using narrowbeam transmitter (TX) and receiver (RX) horn antennas that were boresight-aligned with a test material between the antennas. Overall, 21 different locations were measured for 6 different materials such that the same type of material was tested in at least two locations in order to characterize the effect of penetration loss for materials with similar composition. As shown here, attenuation through common materials ranged between 0.8 dB/cm and 9.9 dB/cm for co-polarized antennas, while cross-polarized antennas exhibited similar attenuation for most materials, but up to 23.4 dB/cm of attenuation for others. The penetration loss results presented here are useful for site-specific planning tools that will model indoor mmWave networks, without the need for expensive measurement campaigns. 
\end{abstract}


\section{Introduction}
The rapid deployment of indoor millimeter wave (mmWave) networks will depend on site-specific planning tools, such as ray-tracers~\cite{Skidmore96a,Tran92a}, to supplement spatial and temporal channel models with site-specific information, including attenuation through common building materials~\cite{Mac15b,Rap15b}. Numerous measurements and analyses were conducted at sub-6 GHz frequencies for modeling indoor environments, and were used to create reliable penetration loss, partition loss, and floor attenuation-based models for network simulations~\cite{Seidel91a,Seidel91b,Seidel92a,Tran92a,Ho93a,Anderson02a,Anderson04a,Durgin98a,Skidmore96a,Panjwani96a,Hawbaker90a,Blankenship97a}.

Penetration loss models have been used to predict attenuation through objects, into buildings, and between floors for indoor and outdoor-to-indoor propagation scenarios~\cite{Rap91c,Zhang94a,Skidmore96a,Durgin98a,Durgin98b,Durgin98d,Anderson02a,Rap13a,Zhao13,Isa15}. Sandhu and Rappaport conducted a literature survey of building penetration loss for below 6 GHz in~\cite{Rap94a}, and Skidmore and Rappaport developed a popular indoor modeling tool called SMT plus~\cite{Skidmore96a}, which eventually led to a commercial product known as SitePlanner and LAN Planner~\cite{Rap04b,ABJ05a}. Zhang and Hwang studied the characteristics of two types of interior walls over the frequency range of 900 MHz to 18 GHz and showed that penetration loss does not necessarily increase linearly or monotonically with respect to frequency~\cite{Zhang94a}. Attenuation for vertically-polarized waves increased non-linearly and non-monotonically with respect to frequency, while horizontally-polarized waves increased monotonically with frequency. Common materials measured from 30 GHz to 50 GHz with different polarized antenna configurations showed that attenuation through a concrete slab has loss of approximately 4.5 dB/cm for both horizontal-to-horizontal (H-H) and vertical-to-vertical (V-V) antenna configurations~\cite{Isa15}. The V-V antenna configuration penetration loss for solid wood in~\cite{Isa15} was reported as 4.19 dB/cm compared to 2.42 dB/cm for the H-H antenna configuration, an approximate 1.8 dB/cm difference due to different polarizations. Zhao \textit{et al.} conducted penetration loss measurements at 28 GHz and found average attenuation through clear glass to be between 3.6 dB and 3.9 dB for different buildings, and found tinted glass induces much more attenuation: 24.5 dB to 40 dB~\cite{Zhao13}.

Floor attenuation factor (FAF) models have been used to predict path loss between locations on different levels or floors of a multi-floored building based on an FAF parameter that accounts for the number of floors separating a transmitter (TX) and receiver (RX), and the building composition~\cite{Seidel91a,Seidel91b,Seidel92a,Seidel92b,Ho93a,Seidel94a,Skidmore96a,Lott01a,Rap04b,ABJ05a,Nguyen16b}. A measurement study in an underground garage (4 levels and $\sim$12 meters deep) at 800 MHz and 2000 MHz showed a frequency-independent FAF of 5.2 dB/m~\cite{Nguyen16b}. FAF values (not normalized to a 1 m reference distance) in an office building at 914 MHz yielded average FAF values that were not a linear function of the number of floors separating the TX and RX, but did increase monotonically~\cite{Seidel91a,Seidel91b,Seidel92a}. Work by Seidel and Rappaport measured an office building with 21 locations at 914 MHz on each floor, and found the FAF was 16.2 dB through the first floor, 27.5 dB through two floors, and 31.6 dB through three floors, indicating that attenuation reduced as the number of floors increased. Measurements in a second office building~\cite{Seidel91a,Seidel91b,Seidel92a} indicated the same trend, with 12.9 dB attenuation loss through one floor, 18.7 dB loss through two floors, and 24.4 dB and 27.0 dB loss through three and four floors, respectively, which indicated a monotonic but non-linear increase in attenuation that tapered off as the number of floors increased. Similar to the FAF model, other models use a combination of the mean large-scale path loss in a line-of-sight (LOS) outdoor environment and an aggregate penetration loss (APL) factor that accounts for exterior wall penetration for an outdoor-to-indoor scenario~\cite{Chamchoy05a,Durgin98a}.

A simple and accurate parabolic building penetration loss (BPL) model for low-loss and high-loss buildings was presented in~\cite{Haneda16a}, and simulations up to 100 GHz were shown to compare well with measurements using the parabolic model (See Eq. (1) in~\cite{Haneda16a}). When using widebeam antennas, observations indicated that loss increased greatly when multiple materials were penetrated. Other measurements showed that attenuation through some materials, such as clear glass, does not necessarily increase with respect to frequency in the mmWave band, and may decrease by as much as 2.8 dB with respect to lower frequencies~\cite{Anderson02a}. Furthermore, interior walls induced 5-6 dB of attenuation on average across measurements of several frequency bands such as 2.5, 5.85, and 60 GHz~\cite{Anderson02a,Durgin98a,Durgin97b}. A study in the terahertz band from 100 GHz to 10 THz reported measured average loss per unit distance for plastic, paper, hardboard, and glass which had losses of 12.47 dB/cm, 15.82 dB/cm, 24.47 dB/cm, and 35.99 dB/cm, respectively~\cite{Kokkoniemi16a}. 

In this paper, penetration loss for six building materials was studied for two antenna polarization configurations at 73 GHz, with the measurement setup presented in Section II. Analysis of the penetration loss measurements is described in Section III, and average penetration loss and normalized average attenuation values with their respective standard deviations are given in Section IV. Conclusions and key observations are provided in Section V.

\begin{figure*}[t!]
\centering
	\includegraphics[width=\textwidth]{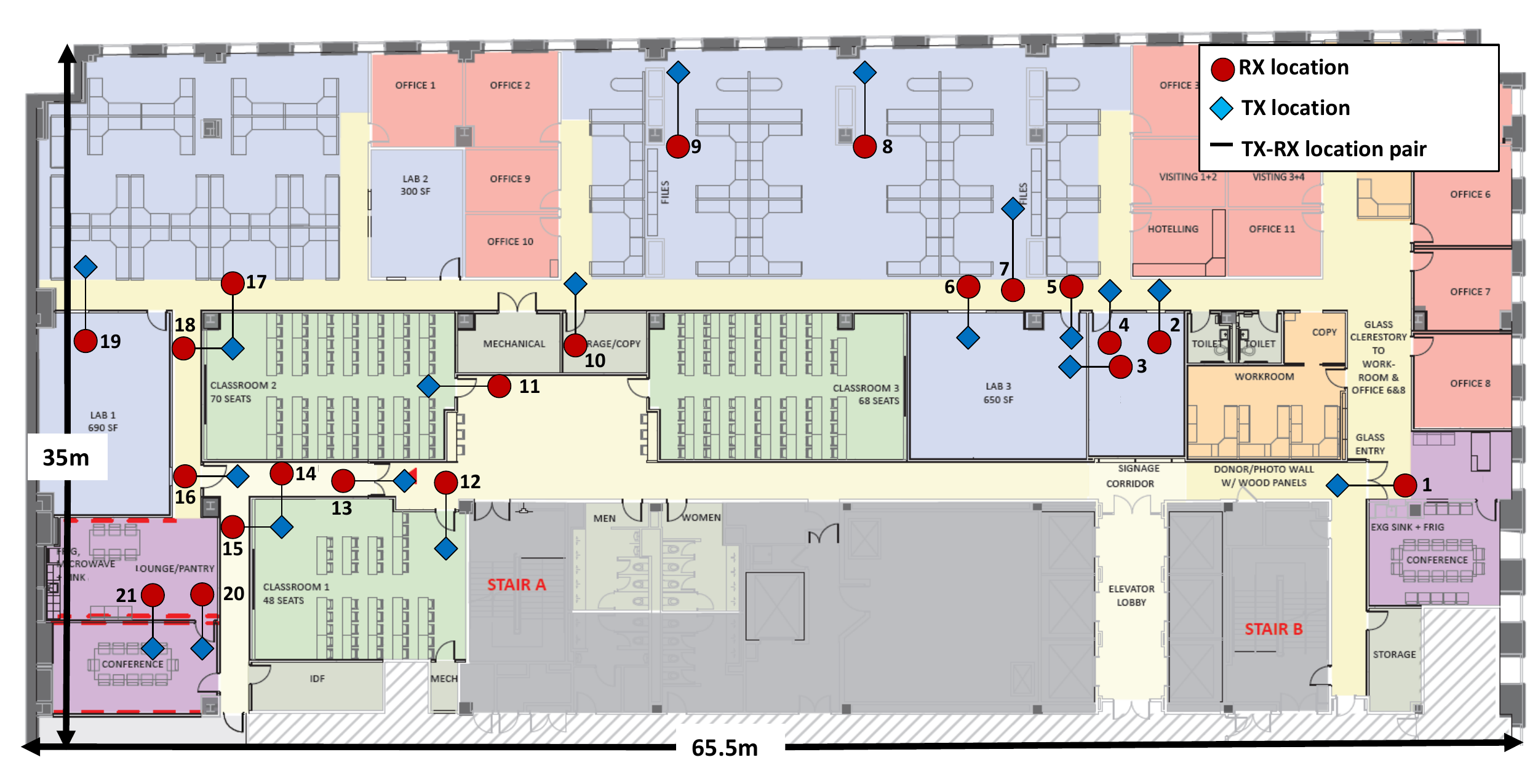}
	\caption{Map of the 9\textsuperscript{th} floor of 2 Metrotech Center, with TX and RX location pairs for the penetration loss measurements. A total of 21 TX/RX location pairs were measured where blue diamonds represent TX locations and red circles represent RX locations. Note that all doors were closed during measurements.}
	\label{fig:MapAll}
\end{figure*}

\section{Measurement Setup}
\subsection{Measurement Hardware}
A wideband sliding correlator channel sounder with a superheterodyne architecture and directional, steerable horn antennas was used to conduct the 73 GHz penetration loss measurements~\cite{Mac17a,Mac17c,Mac14a}. At the transmitter a pseudorandom-noise (PN) code was generated at baseband at 500 Megachips-per-second (Mcps) and was subsequently modulated with a 5.625 GHz intermediate frequency (IF). The wideband IF signal was mixed with a 67.875 GHz local oscillator (LO) to obtain a center frequency of 73.5 GHz with a 1 GHz radio frequency (RF) null-to-null bandwidth that was transmitted through a horn antenna. An identical superheterodyne architecture was employed at the RX, with a sliding correlator for baseband processing which resulted in the acquisition of power delay profiles (PDPs). The TX and RX antennas were identical with 20 dBi of gain and 15 degree azimuth and elevation half-power beamwidth (HPBW). The antennas were always boresight-aligned with the material under test (MUT) placed in between the TX and RX antennas. Channel sounder specifications are provided in Table~\ref{spec_tbl} and further descriptions of the measurement system can be found in~\cite{Mac15b,Mac17a,Mac17c}.

\begin{table}[h]
	\centering
	\caption{Channel sounder specifications for penetration loss measurements.}
	\label{spec_tbl}
	\begin{center}
	\begin{tabu}{|c|[1.6pt] p{2.5cm}|}
	\hline 
	\textbf{Carrier Frequency} 			& \centering{\textbf{73.5 GHz}}	\\ \specialrule{1.5pt}{0pt}{0pt}
	\textbf{TX PN Code Chip Rate} 		&	\centering{500 Mcps} 		\\ \hline
    \textbf{TX/RX IF Frequency}			&	\centering{5.625 GHz}			\\ \hline
    \textbf{TX/RX LO Frequency}			&	\centering{67.875 GHz}		\\ \hline
    \textbf{RF Bandwidth (Null-to-Null)}&	\centering{1 GHz}			\\ \hline
	\textbf{Max. TX Output Power} 		&	\centering{14.1 dBm}		\\ \hline
	\textbf{TX/RX Antenna Gain} 		&	\centering{20 dBi}			\\ \hline
	\textbf{TX and RX Azimuth/Elevation HPBW} 		&	\centering{15$^{\circ}$/15$^{\circ}$}	\\ \hline
	\textbf{TX and RX Antenna Height} 		&	\centering{1.5 m}			\\ \hline 
	\textbf{Multipath Time Resolution}	&	\centering{2 ns}			\\ \hline
	\textbf{TX Polarization} 			&	\centering{Vertical}		\\ \hline
	\textbf{RX Polarization} 			&	\centering{Vertical / Horizontal} \\ \hline
	\end{tabu}
	\end{center}
\end{table}

\subsection{Measurement Environment}
\begin{figure}[t]
	\centering
	\includegraphics[width=\columnwidth]{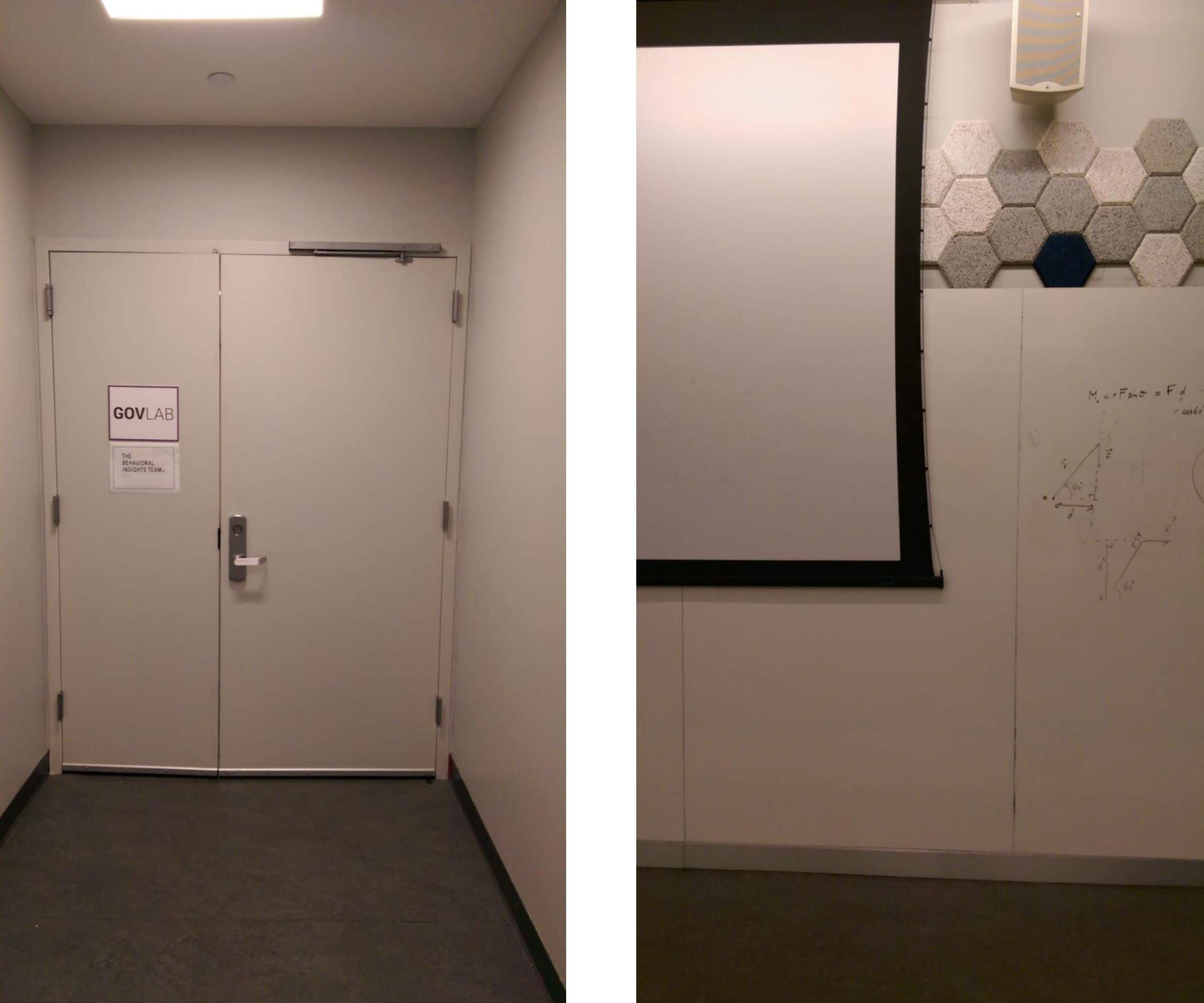}
    \caption{Examples pictures of two different material types, a steel door (\textit{left}) and a whiteboard writing wall (\textit{right}). The whiteboard writing wall had a projector screen covering a portion of the surface as shown in the picture, however penetration through the screen was not measured.}\label{fig:Materials}
\end{figure}

Measurements were conducted in the NYU WIRELESS research center, located on the 9$^{\textrm{th}}$ floor of 2 Metrotech Center, an office building in Brooklyn, New York, constructed in 1989. The building resembles a typical office environment, including interior walls made of drywall with metal studs, whiteboard writing walls, clear glass windows, glass doors, closet doors, and steel doors. A total of 210 PDPs of received power were recorded at 21 TX/RX location pairs as shown in Fig.~\ref{fig:MapAll} and in Table~\ref{penresult_tbl}. Interior plasterboard walls (locations 3, 14, 17, and 21) had an average thickness of 13.7 centimeters (cm) and did not contain a metal stud layer in any of the tested locations. Whiteboard writing walls (locations 15 and 18) were considered to be interior walls with an additional layer composed of a whiteboard made from fiberboard, and had an average thickness of 21.4 cm. Clear glass windows (locations 2, 6, and 19) were approximately 50 cm by 50 cm with a thickness of 1 cm. The glass doors (locations 1, 4, 5, 11, and 12) had an 80 cm wide metal frame with a 50 cm wide inset glass panel, which had a thickness of 1 cm. Closet doors (locations 7, 8, and 9) were composed of a 7 cm thick solid layer of medium-density fiberboard (MDF) and had a width of approximately 45 cm. Steel doors (locations 10, 13, 16, and 20) had an average thickness of 5.3 cm, and contained two metal layers with a hollow interior. Two examples of the measured building materials are shown in Fig.~\ref{fig:Materials}.

\begin{figure}[t]
	\centering
	\includegraphics[width=\columnwidth]{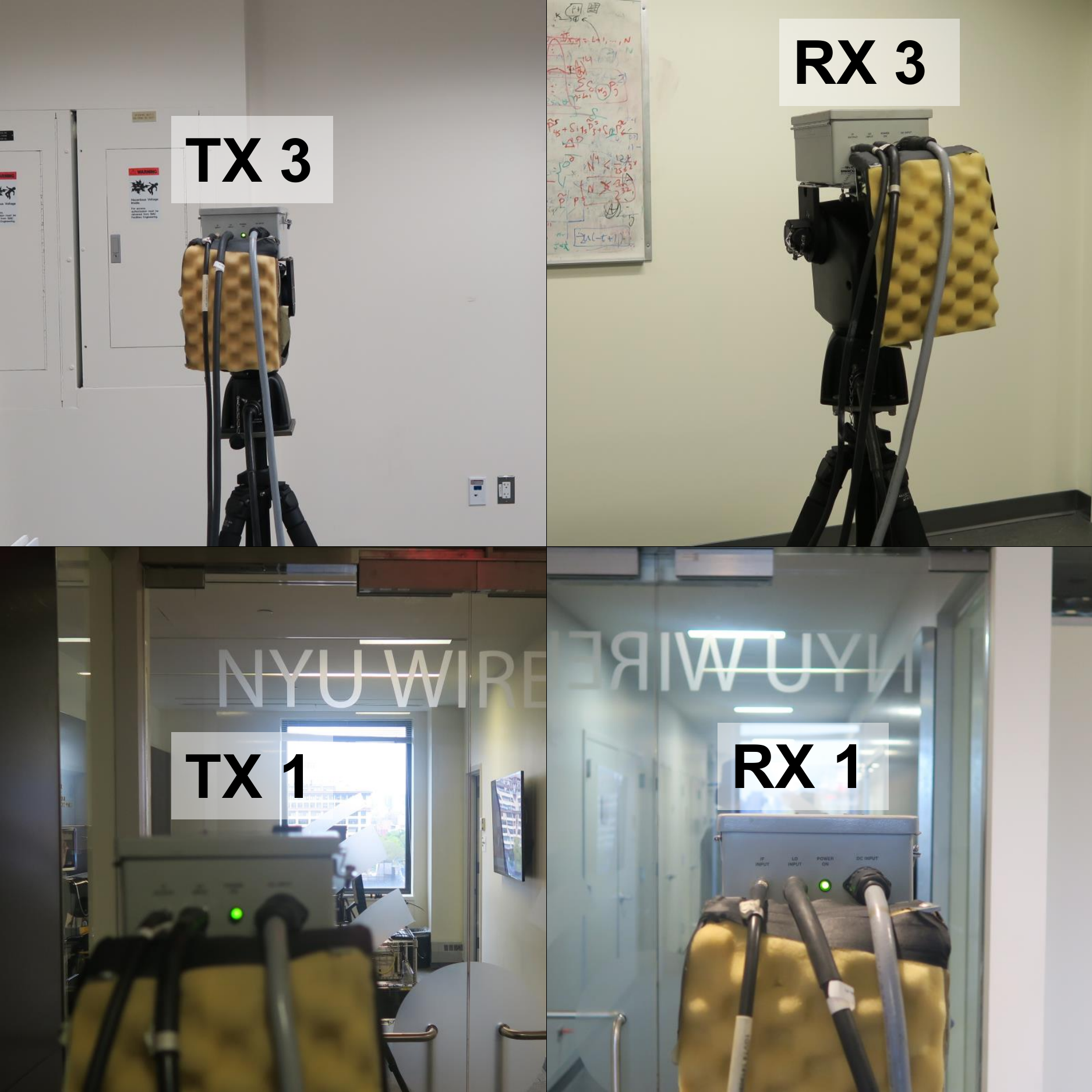}
    \caption{Measurement setup at two locations: Loc. 1 and Loc. 3. The TX and RX antennas were boresight-aligned and placed at most 1.5 m away from either side of the surface of the MUT. TX 1 and RX 1 tested a glass door; TX 3 and RX 3 tested a plasterboard wall.}\label{fig:Setup}
\end{figure}

\begin{figure}[h]
	\centering
    \includegraphics[width=\columnwidth]{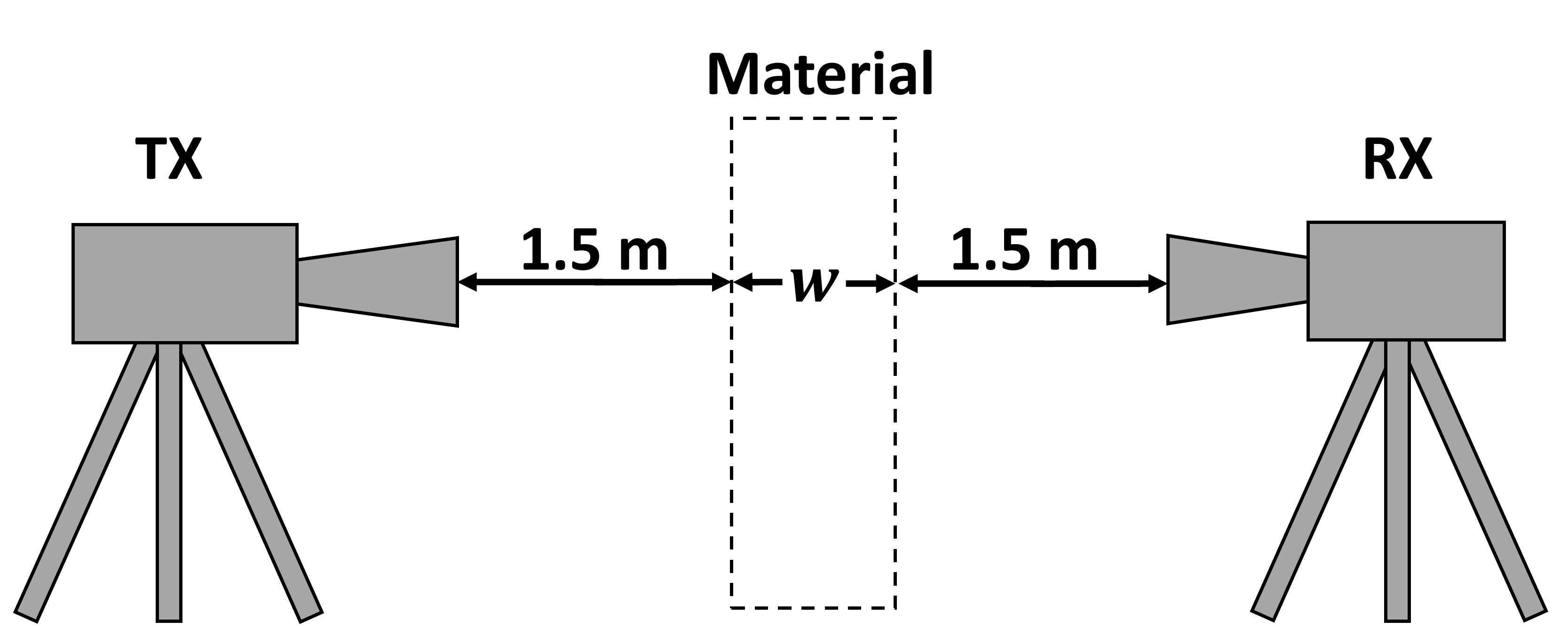}
    \caption{Profile diagram of the measurement setup where the TX and RX antennas were placed at most 1.5 m from either side of the MUT, which had a width of $w$.}\label{fig:diagram}
\end{figure}

\subsection{Measurement of Received Power}
At each location, the TX and RX were both placed at 1.5 m from either side of the surface of the MUT, with the TX and RX antennas boresight-aligned, each at a height of 1.5 m. The distance was chosen to ensure the MUT was in the far field of the antennas while minimizing the width of the spread of the transmitted wave upon the material, which was 40 cm x 40 cm at a distance of 1.5 m. However, at three locations (15, 16, and 18), the corridors in the building were too narrow to allow the RX to be placed at a 1.5 m distance from the wall being measured, and in these cases the RX was placed about 1 m from the material. Ten PDPs were captured at each location with the TX and RX antennas boresight-aligned, with 5 PDPs recorded for both V-V and V-H antenna configurations. The measurement setup remained fixed for each set of 5 repeated PDPs for consistency, averaging purposes, and to diminish any movement in the channel caused by humans. Two examples of the setup of the TX and RX at locations 1 and 3 are shown in Fig.~\ref{fig:Setup}, and a diagram of the setup is depicted in Fig.~\ref{fig:diagram}. The measured received power was determined as the power in the first-arriving multipath component (MPC)~\cite{Durgin98e,Newhall97a}, rather than the spatial average or integration over all MPCs, as used in some models of wall penetration~\cite{A5GCM15}.  The first-arriving path was assumed to be the penetrating path, whereas later-arriving paths were ignored as they were considered to come from reflections of the signal from nearby walls and other obstructions in the environment~\cite{Joshi05a}.
 
\section{Propagation Analysis}
\subsection{Penetration Loss Calculation}
Penetration loss due to a single material was determined as the difference between measured average received power in the first arriving MPC and the received power in theoretical free space with the same T-R separation distance, including the width of the material ($w$). The penetration loss, $L$, of a material was obtained by:
\begin{equation}\label{eq:penloss_eq}
L[\dB] = P_{r, \FS} - P_{r, \text{meas.}}
\end{equation}
\begin{equation}\label{eq:Friis}
P_{r, \FS}[\dBm] = P_t + G_t + G_r + 20\log_{10}\left(\frac{c}{4\pi d f_c}\right)
\end{equation}
where $d$ is the T-R separation distance, which was counted as the sum of the distance between the TX and RX antennas and the width ($w$) of the MUT. $P_t$ is the transmit power in dBm, $G_t$ is the TX antenna gain in dBi, $G_r$ is the RX antenna gain in dBi, $c$ is the speed of light in free space, $f_c$ is the carrier frequency, $P_{r, \text{meas}}$ is the measured received power of the first MPC of a single PDP, and $P_{r, \FS}$ is the theoretical received power in free space (FS) obtained using Friis' free space transmission formula in~\eqref{eq:Friis}.

Average penetration losses at each TX-RX location combination were computed as the linear average in milliwatts (mW) of the 5 PDPs recorded for V-V and V-H antenna polarization combinations, respectively. For V-H measurements, an additional cross-polarization discrimination (XPD) factor was subtracted from the received power to remove the cross-polarized antenna mismatch and to isolate penetration loss for the MUT. The XPD factor was determined empirically from five free space measurements of received power for V-V antennas compared to the received power for V-H antennas for distances ranging from 2.6 m to 3.0 m, in 0.1 m increments. The XPD factor for each distance was then calculated by subtracting the V-H received power from the V-V received power. All five distances measured had an XPD within 1.5 dB of each other with no dependence on distance, which demonstrated that all measurements, even with the RX at 0.86 m from the MUT for one case, were in the far-field of the TX antenna. The measurement system XPD factor of 27.1 dB was calculated through averaging in linear over all five measurement distances.

\subsection{Penetration Loss Material Types}
\textcolor{black}{Data were grouped according to the MUT between the TX and RX antennas. Average penetration losses were determined by averaging the penetration loss (in linear) for data of similar materials and for each polarization and then converting back to log scale in dB. Average penetration loss standard deviations were calculated with the penetration loss values (in dB scale) for data of a common MUT. Average penetration loss calculated in log scale was found to be within 0.2 dB of averaging in linear scale. The average penetration loss for each material type does not represent the measured penetration loss at any specific location, but rather an aggregate penetration loss of a typical material which might be found in an office environment. It is expected that the average penetration loss for each material type will be similar for other office buildings~\cite{Seidel91a,Seidel91b,Seidel92a, Zhao13}.}

\textcolor{black}{For simple comparison purposes, the average penetration loss for each material type (in dB scale) was divided by the average width of the group of common test materials to determine the normalized average penetration loss in dB/cm. The standard deviation, $\sigma_N$, for normalized average penetration loss was obtained by taking the standard deviation of the normalized penetrations losses in dB/cm among the data in each group of common materials.}

\section{Results}
A summary of all the average penetration losses through each material type for V-V and V-H antenna polarizations is provided in Table~\ref{penresult_tbl} along with the standard deviation of penetration loss for all locations which tested a common material type and is given by $\sigma_L$ in dB. The XPD factor of 27.1 dB was subtracted from the cross-polarization measurements to provide the V-H penetration loss values in Table~\ref{penresult_tbl}. Normalized average penetration loss values with their respective standard deviations ($\sigma_N$ in dB/cm) are also provided in Table~\ref{penresult_tbl}. The highest average penetration loss was 73.8 dB, which occurred through whiteboard writing walls for V-V polarized antennas, however, the normalized loss for this material was 3.5 dB/cm, about a 3 dB/cm increase over the normalized average attenuation for typical walls. The whiteboard writing walls were the thickest materials tested, ranging in width from 21.1 cm to 21.6 cm, with the whiteboard component accounting for 5.5 cm of the overall thickness. The whiteboard component was composed of two layers of dense fiberboard, which in consideration of the increase in normalized attenuation loss for whiteboard writing walls as compared with normal walls, likely contributed to a majority of the attenuation through the material. The lowest penetration loss measured for any material was 5.1 dB for glass doors with V-V polarized antennas, and a normalized average attenuation of 5.1 dB/cm since the thickness of the glass material in the door was 1 cm for all locations. Clear glass windows had 2 dB greater average penetration loss than glass doors for V-V polarized antennas, likely due to the antenna spread upon the material of 1 m which was wider than the window itself at all of the clear glass window locations. A portion of the transmitted energy may have reflected and scattered off of the metal frame surrounding the window, thus reducing the received power in the first-arriving MPC as compared with clear glass doors that contained larger glass panels.

\begin{table}[t!]
	\centering
	\renewcommand{\arraystretch}{1.3}
	\caption{Average penetration loss values with respective standard deviations. ``Pol." represents the TX-RX antenna polarization configuration, ``No. of Loc." is the total number of measured locations for a material type, ``Loss" is the average penetration loss in dB, $\sigma_L$ is the standard deviation of the penetration loss over all materials of the same type in dB. ``Norm. Avg. Atten." is the normalized average attenuation in dB/cm, and $\sigma_{N}$ is the standard deviation of the normalized average attenuation over all materials of the same type in dB/cm. The XPD factor of 27.1 dB was subtracted from the cross-polarized antenna measurements to calculate the  V-H penetration loss values.}\label{penresult_tbl}
	\fontsize{9}{9}\selectfont
	\begin{tabular}{|>{\centering\arraybackslash}m{1.5cm}|>{\centering\arraybackslash}m{0.6cm}|>{\centering\arraybackslash}m{0.6cm}?>{\centering\arraybackslash}m{0.6cm}|>{\centering\arraybackslash}m{0.6cm}|>{\centering\arraybackslash}m{1.0cm}|>{\centering\arraybackslash}m{1.2cm}|} \hline
		\multicolumn{7}{|c|}{\textbf{Penetration Loss of Common Building}} \\
		\multicolumn{7}{|c|}{\textbf{Materials for 73 GHz V-V and V-H}} \\ \hline
		\textbf{Material} &	\textbf{No. of Loc.} &	\textbf{Pol.} &	\textbf{Loss (dB)} &	\textbf{$\bm{\sigma_L}$ (dB)} &	\textbf{Norm. Avg. Atten. (dB/cm)} &	\textbf{$\bm{\sigma_{N}}$ (dB/cm)}	\\ \specialrule{1.5pt}{0pt}{0pt}
		\multirow{2}{*}{Glass Door}	&	\multirow{2}{*}{5} &	V-V	& 5.1 &	1.2 &	5.1 & 1.2  \\ \cline{3-7}
		&	&	V-H & 	23.4 &	7.1 &	23.4 &	7.1  \\ \hline
		\multirow{2}{*}{Clear Glass} &	\multirow{2}{*}{3} &	V-V &	7.1 &	2.3 &	7.1 &	2.3  \\ \cline{3-7}
		&	&	V-H &	18.3 &	3.4 &	18.3 &	3.4  \\ \hline
		\multirow{2}{*}{Wall} &	\multirow{2}{*}{4} &	V-V &	10.6 &	5.6 &	0.8 &	0.3  \\ \cline{3-7}
		&	&	V-H &	11.7 &	6.2 &	0.8 &	0.4  \\ \hline
		\multirow{2}{*}{Closet Door} &	\multirow{2}{*}{3} &	V-V &	32.3 &	8.2 &	4.6 &	1.2  \\ \cline{3-7}
		&	&	V-H &	16.3 &	4.2 &	2.3 &	0.6  \\ \hline
		\multirow{2}{*}{Steel Door} &	\multirow{2}{*}{4} &	V-V	& 52.2 &	4.0 &	9.9 &	0.9  \\ \cline{3-7}
		&	&	V-H	&	48.3 &	5.6 &	9.2 &	0.5  \\ \hline
		\multirow{2}{*}{\begin{tabular}[x]{@{}c@{}}Whiteboard \\ W. Wall \end{tabular}} &	\multirow{2}{*}{2} &	V-V &	73.8 &	9.8 &	3.5 &	0.5  \\ \cline{3-7}
		&	&	V-H &	58.1 &	3.0 &	2.7 &	0.2  \\ \hline
		
	\end{tabular}
\end{table}

Penetration loss increased for glass doors, walls, and clear glass windows for V-H polarization as compared to the V-V antenna penetration losses. Walls were observed to have 10.6 dB and 11.7 dB average penetration loss at 73 GHz for co- and cross-polarized antenna measurements, respectively, which is higher than penetration losses of 5 to 6 dB reported in other studies at lower frequencies with similar materials~\cite{Anderson02a,Andersen04a,Durgin98a}. The increase in attenuation caused by walls at 73 GHz may be due to differences in the wall material specifications; the walls tested in this paper were up to 16 cm thick and may have contained insulating or reflective material layers inside. For steel doors, closet doors, and whiteboard writing walls, the average V-H penetration loss decreased as compared to V-V penetration, specifically by up to 16 dB in the case of closet doors. The materials for which penetration loss decreased from V-V to V-H measurements contained the most reflective materials (whiteboard, metal, and particle board), which suggests that the transmitted wave may have depolarized as it propagated through different layers of the material. Average penetration loss did not demonstrate a uniform increase or decrease with respect to co-polarized versus cross-polarized antenna configurations, but attenuation was observed to vary by as much as 18.3 dB (increase) in the case of glass doors and by 16 dB (decrease) in the case of closet doors.

The whiteboard writing wall locations had a penetration loss standard deviation $\sigma_L$ of 9.8 dB for V-V, making it the highest observed standard deviation over all the materials. For V-H, glass doors had the highest standard deviation for average penetration loss with a value of $\sigma_L=7.1$ dB. Over all materials, the standard deviation of the average penetration loss varied from 1.2 dB to 9.8 dB. \textcolor{black}{The standard deviation of the normalized average penetration loss ($\sigma_N$) was typically less than the standard deviation of the average penetration loss by approximately 1 to 5 dB. The values for $\sigma_{N}$ contain information about variation in the width of the material between locations as well as the penetration loss. Therefore, the normalized penetration loss standard deviations were the same or less than the average penetration loss standard deviations because the width of a material at a specific location was typically within 0 to 1 cm thickness of the same materials from other locations, and is the case for all materials.} 

While penetration loss was determined to be 10.6 dB on average for typical walls, whiteboard writing walls were observed to have over 60 dB greater attenuation. Furthermore, the attenuation through two other thick materials (steel doors and closet doors) was found to range from 30 dB to 50 dB. Such large penetration losses suggest that certain common building materials will create barriers unsuitable for penetration-based propagation in 73 GHz systems. The penetration loss values presented in this paper could be embedded into 3D ray-tracer simulations or indoor planning tools for mmWave indoor propagation modeling, similar to studies and tools at lower frequencies~\cite{Isa15,Seidel91a,Seidel91b,Seidel92a,Schaubach92a,Tran92a,Ho93a,Seidel94a,Skidmore96a}.

\section{Conclusion}
Penetration loss measurements at 73 GHz for V-V and V-H polarization configurations were presented for six common building materials found in indoor office environments such as glass doors, glass windows, steel doors, MDF closet doors, drywall, and whiteboards. Attenuation was found to increase by up to 18.3 dB between V-V and V-H measurements of glass doors, however, penetration loss was not necessarily found to increase or decrease based strictly on the antenna polarization configuration. Measurements of closet doors showed a 16 dB decrease in average penetration loss for cross-polarized antennas as compared with co-polarized antennas. The standard deviation observed between measurements of the same material at different locations was as large as 9.8 dB in the case of whiteboard writing walls for V-V, and the lowest standard deviation was observed for V-V glass door measurements, with a value of 1.2 dB. The standard deviations of the penetration loss measurements from Table~\ref{penresult_tbl} were found to be higher than results at frequencies below 6 GHz~\cite{Durgin97b,Anderson02a,Anderson04a}. Whiteboard writing walls had an average penetration loss of 73.8 dB for V-V, but only 3.5 dB/cm normalized average attenuation, which is comparable to a closet door made from MDF, which had a normalized average attenuation of 4.6 dB/cm for V-V antennas. Co-polarized penetration loss for glass doors and windows was found to be 5 to 7 dB, which suggests that 73 GHz systems could be suitable for propagation between partitions with large glass windows or glass walls. On the other hand, the large penetration losses observed for metal doors and thick whiteboard walls may be useful for interference isolation between neighboring rooms made of these materials. The penetration losses provided in Table~\ref{penresult_tbl} and in this paper may be used in mmWave ray-tracers/site-specific tools used for estimating average path loss in indoor office environments~\cite{Seidel91a,Seidel91b,Seidel92a,Ho93a,Seidel94a,Schaubach92a,Tran92a,Skidmore96a}.

\bibliography{MacCartney_Bibv5}

\begin{thebibliography}{10}
\providecommand{\url}[1]{#1}
\csname url@samestyle\endcsname
\providecommand{\newblock}{\relax}
\providecommand{\bibinfo}[2]{#2}
\providecommand{\BIBentrySTDinterwordspacing}{\spaceskip=0pt\relax}
\providecommand{\BIBentryALTinterwordstretchfactor}{4}
\providecommand{\BIBentryALTinterwordspacing}{\spaceskip=\fontdimen2\font plus
\BIBentryALTinterwordstretchfactor\fontdimen3\font minus
  \fontdimen4\font\relax}
\providecommand{\BIBforeignlanguage}[2]{{%
\expandafter\ifx\csname l@#1\endcsname\relax
\typeout{** WARNING: IEEEtran.bst: No hyphenation pattern has been}%
\typeout{** loaded for the language `#1'. Using the pattern for}%
\typeout{** the default language instead.}%
\else
\language=\csname l@#1\endcsname
\fi
#2}}
\providecommand{\BIBdecl}{\relax}
\BIBdecl

\bibitem{Skidmore96a}
R.~R. Skidmore, T.~S. Rappaport, and A.~L. Abbott, ``Interactive coverage
  region and system design simulation for wireless communication systems in
  multifloored indoor environments: {SMT PLUS},'' in \emph{Proceedings of the
  5th IEEE International Conference on Universal Personal Communications},
  vol.~2, Sept. 1996, pp. 646--650.

\bibitem{Tran92a}
T.~T. Tran and T.~S. Rappaport, ``Site specific propagation prediction models
  for {PCS} design and installation,'' in \emph{MILCOM 92 Conference Record},
  vol.~3, Oct. 1992, pp. 1062--1065.

\bibitem{Mac15b}
G.~R. {MacCartney, Jr.} \emph{et~al.}, ``Indoor office wideband millimeter-wave
  propagation measurements and models at 28 {GHz} and 73 {GHz} for ultra-dense
  {5G} wireless networks ({Invited Paper}),'' \emph{IEEE Access}, vol.~3, pp.
  2388--2424, Oct. 2015.

\bibitem{Rap15b}
T.~S. Rappaport \emph{et~al.}, ``Wideband millimeter-wave propagation
  measurements and channel models for future wireless communication system
  design ({Invited Paper}),'' \emph{IEEE Transactions on Communications},
  vol.~63, no.~9, pp. 3029--3056, Sept. 2015.

\bibitem{Seidel91a}
S.~Y. Seidel and T.~S. Rappaport, ``900 {MHz} path loss measurements and
  prediction techniques for in-building communication system design,'' in
  \emph{1991 Proceedings of the 41st IEEE Vehicular Technology Conference}, May
  1991, pp. 613--618.

\bibitem{Seidel91b}
------, ``Path loss prediction in multifloored buildings at 914 {MHz},''
  \emph{Electronics Letters}, vol.~27, no.~15, pp. 1384--1387, July 1991.

\bibitem{Seidel92a}
------, ``914 {MHz} path loss prediction models for indoor wireless
  communications in multifloored buildings,'' \emph{IEEE Transactions on
  Antennas and Propagation}, vol.~40, no.~2, pp. 207--217, Feb. 1992.

\bibitem{Ho93a}
C.~M.~P. Ho and T.~S. Rappaport, ``Wireless channel prediction in a modern
  office building using an image-based ray tracing method,'' in \emph{1993 IEEE
  Global Telecommunications Conference (GLOBECOM '93), including a
  Communications Theory Mini-Conference}, vol.~2, Nov. 1993, pp. 1247--1251.

\bibitem{Anderson02a}
C.~R. Anderson \emph{et~al.}, ``In-building wideband multipath characteristics
  at 2.5 and 60 {GHz},'' in \emph{Proceedings IEEE 56th Vehicular Technology
  Conference}, vol.~1, Sept. 2002, pp. 97--101.

\bibitem{Anderson04a}
C.~R. Anderson and T.~S. Rappaport, ``In-building wideband partition loss
  measurements at 2.5 and 60 {GHz},'' \emph{IEEE Transactions on Wireless
  Communications}, vol.~3, no.~3, pp. 922--928, May 2004.

\bibitem{Durgin98a}
G.~D. Durgin, T.~S. Rappaport, and H.~Xu, ``Measurements and models for radio
  path loss and penetration loss in and around homes and trees at 5.85 {GH}z,''
  \emph{IEEE Transactions on Communications}, vol.~46, no.~11, pp. 1484--1496,
  Nov. 1998.

\bibitem{Panjwani96a}
M.~A. Panjwani, A.~L. Abbott, and T.~S. Rappaport, ``Interactive computation of
  coverage regions for wireless communication in multifloored indoor
  environments,'' \emph{IEEE Journal on Selected Areas in Communications},
  vol.~14, no.~3, pp. 420--430, Apr. 1996.

\bibitem{Hawbaker90a}
D.~A. Hawbaker and T.~S. Rappaport, ``Indoor wideband radiowave propagation
  measurements at 1.3 {GHz} and 4.0 {GHz},'' \emph{Electronics Letters},
  vol.~26, no.~21, pp. 1800--1802, Oct. 1990.

\bibitem{Blankenship97a}
T.~K. Blankenship, D.~M. Kriztman, and T.~S. Rappaport, ``Measurements and
  simulation of radio frequency impulsive noise in hospitals and clinics,'' in
  \emph{1997 IEEE 47th Vehicular Technology Conference. Technology in Motion},
  vol.~3, May 1997, pp. 1942--1946.

\bibitem{Rap91c}
T.~S. Rappaport, ``Wireless personal communications: trends and challenges,''
  \emph{IEEE Antennas and Propagation Magazine}, vol.~33, no.~5, pp. 19--29,
  Oct. 1991.

\bibitem{Zhang94a}
Y.~P. Zhang and Y.~Hwang, ``Measurements of the characteristics of indoor
  penetration loss,'' in \emph{1994 IEEE 44th Vehicular Technology Conference
  (VTC)}, vol.~3, June 1994, pp. 1741--1744.

\bibitem{Durgin98b}
G.~D. Durgin, T.~S. Rappaport, and H.~Xu, ``Partition-based path loss analysis
  for in-home and residential areas at 5.85 {GH}z,'' in \emph{1998 IEEE Global
  Communications Conference (GLOBECOM)}, vol.~2, Nov. 1998, pp. 904--909.

\bibitem{Durgin98d}
G.~Durgin, T.~S. Rappaport, and H.~Xu, ``5.85-{GHz} radio path loss and
  penetration loss measurements in and around homes and trees,'' \emph{IEEE
  Communications Letters}, vol.~2, no.~3, pp. 70--72, Mar. 1998.

\bibitem{Rap13a}
T.~S. Rappaport \emph{et~al.}, ``{Millimeter Wave Mobile Communications for
  {5G} Cellular: It Will Work!}'' \emph{IEEE Access}, vol.~1, pp. 335--349, May
  2013.

\bibitem{Zhao13}
H.~Zhao \emph{et~al.}, ``28 {GH}z millimeter wave cellular communication
  measurements for reflection and penetration loss in and around buildings in
  {New York city},'' in \emph{2013 IEEE International Conference on
  Communications (ICC)}, June 2013, pp. 5163--5167.

\bibitem{Isa15}
A.~K.~M. Isa, A.~Nix, and G.~Hilton, ``Impact of diffraction and attenuation
  for material characterisation in millimetre wave bands,'' in \emph{2015
  Loughborough Antennas and Propagation Conference (LAPC)}, Nov. 2015, pp.
  1--4.

\bibitem{Rap94a}
T.~S. Rappaport and S.~Sandhu, ``Radio-wave propagation for emerging wireless
  personal-communication systems,'' \emph{IEEE Antennas and Propagation
  Magazine}, vol.~36, no.~5, pp. 14--24, Oct. 1994.

\bibitem{Rap04b}
\BIBentryALTinterwordspacing
T.~S. Rappaport and R.~Skidmore, ``System and method for ray tracing using
  reception surfaces,'' Dec. 2004, {US Patent 10/830,445}. [Online]. Available:
  \url{https://www.google.com/patents/US20040259554}
\BIBentrySTDinterwordspacing

\bibitem{ABJ05a}
\BIBentryALTinterwordspacing
{Austin Business Journal}, ``Motorola buys wireless valley,'' Dec. 2005.
  [Online]. Available:
  \url{http://www.bizjournals.com/austin/stories/2005/12/19/daily46.html}
\BIBentrySTDinterwordspacing

\bibitem{Seidel92b}
S.~Y. Seidel and T.~S. Rappaport, ``A ray tracing technique to predict path
  loss and delay spread inside buildings,'' in \emph{1992 IEEE Global
  Communications Conference (GLOBECOM): Communication for Global Users},
  vol.~2, Dec. 1992, pp. 649--653.

\bibitem{Seidel94a}
------, ``Site-specific propagation prediction for wireless in-building
  personal communication system design,'' \emph{IEEE Transactions on Vehicular
  Technology}, vol.~43, no.~4, pp. 879--891, Nov. 1994.

\bibitem{Lott01a}
M.~Lott and I.~Forkel, ``A multi-wall-and-floor model for indoor radio
  propagation,'' in \emph{IEEE VTS 53rd Vehicular Technology Conference, Spring
  2001. Proceedings (Cat. No.01CH37202)}, vol.~1, May 2001, pp. 464--468.

\bibitem{Nguyen16b}
H.~C. Nguyen \emph{et~al.}, ``A simple statistical signal loss model for deep
  underground garage,'' in \emph{2016 IEEE 84th Vehicular Technology Conference
  (VTC2016-Fall)}, Sept. 2016, pp. 1--5.

\bibitem{Chamchoy05a}
M.~Chamchoy, P.~Jaturatussanai, and S.~Promwong, ``Empirically based path loss
  and penetration loss models for {UWB} communication in residential
  environment,'' in \emph{2005 Fifth International Conference on Information,
  Communications and Signal Processing}, Dec. 2005, pp. 278--281.

\bibitem{Haneda16a}
K.~Haneda \emph{et~al.}, ``{5G 3GPP}-like channel models for outdoor urban
  microcellular and macrocellular environments,'' in \emph{2016 IEEE 83rd
  Vehicular Technology Conference (VTC2016-Spring)}, May 2016, pp. 1--7.

\bibitem{Durgin97b}
G.~Durgin, N.~Patwari, and T.~S. Rappaport, ``Improved 3{D} ray launching
  method for wireless propagation prediction,'' \emph{Electronics Letters},
  vol.~33, no.~16, pp. 1412--1413, July 1997.

\bibitem{Kokkoniemi16a}
J.~Kokkoniemi, J.~Lehtomäki, and M.~Juntti, ``Measurements on penetration loss
  in terahertz band,'' in \emph{2016 10th European Conference on Antennas and
  Propagation (EuCAP)}, Apr. 2016, pp. 1--5.

\bibitem{Mac17a}
G.~R. {MacCartney, Jr.} and T.~S. Rappaport, ``A flexible millimeter-wave
  channel sounder with absolute timing,'' \emph{IEEE Journal on Selected Areas
  in Communications}, Aug. 2017.

\bibitem{Mac17c}
G.~R. {MacCartney, Jr.} \emph{et~al.}, ``A flexible wideband millimeter-wave
  channel sounder with local area and {NLOS to LOS} transition measurements,''
  in \emph{2017 IEEE International Conference on Communications (ICC)}, May
  2017, pp. 1--7.

\bibitem{Mac14a}
G.~R. {MacCartney, Jr.} and T.~S. Rappaport, ``73 {GHz} millimeter wave
  propagation measurements for outdoor urban mobile and backhaul communications
  in {New York City},'' in \emph{2014 IEEE International Conference on
  Communications (ICC)}, June 2014, pp. 4862--4867.

\bibitem{Durgin98e}
G.~Durgin and T.~S. Rappaport, ``Basic relationship between multipath angular
  spread and narrowband fading in wireless channels,'' \emph{Electronics
  Letters}, vol.~34, no.~25, pp. 2431--2432, Dec. 1998.

\bibitem{Newhall97a}
W.~G. Newhall and T.~S. Rappaport, ``An antenna pattern measurement technique
  using wideband channel profiles to resolve multipath signal components,'' in
  \emph{Antenna Measurement Techniques Association 19th Annual Meeting \&
  Symposium}, Nov. 1997, pp. 17--21.

\bibitem{A5GCM15}
\BIBentryALTinterwordspacing
{Aalto University, AT\&T, BUPT, CMCC, Ericsson, Huawei, Intel, KT Corporation,
  Nokia, NTT DOCOMO, New York University, Qualcomm, Samsung, University of
  Bristol, and University of Southern California}, ``{5G} channel model for
  bands up to 100 {GHz},'' 2016, Oct. 21. [Online]. Available:
  \url{http://www.5gworkshops.com/5GCM.html}
\BIBentrySTDinterwordspacing

\bibitem{Joshi05a}
G.~G. Joshi \emph{et~al.}, ``Near-ground channel measurements over
  line-of-sight and forested paths,'' \emph{IEEE Proceedings - Microwaves,
  Antennas and Propagation}, vol. 152, no.~6, pp. 589--596, Dec. 2005.

\bibitem{Andersen04a}
J.~B. Andersen \emph{et~al.}, ``A 16 by 32 wideband multichannel sounder at 5
  {GHz for MIMO},'' in \emph{IEEE Antennas and Propagation Society
  International Symposium}, vol.~2, June 2004, pp. 1263--1266.

\bibitem{Schaubach92a}
K.~R. Schaubach, N.~J. Davis, and T.~S. Rappaport, ``A ray tracing method for
  predicting path loss and delay spread in microcellular environments,'' in
  \emph{1992 Proceedings of the Vehicular Technology Society 42nd VTS
  Conference - Frontiers of Technology}, vol.~2, May 1992, pp. 932--935.

\end{thebibliography}
\bibliographystyle{IEEEtran}
\end{document}